\documentstyle[aps,prl,preprint,floats,epsfig]{revtex}

\begin{document}

\preprint{\tighten\vbox{\hbox{\hfil CLEO CONF 00-5}
                        \hbox{\hfil ICHEP00 863}
}}

\title{Evidence of New States Decaying into $\Xi_c^{\prime} \pi$}  

\author{CLEO Collaboration}
\date{\today}

\maketitle
\tighten

\begin{abstract} 
Using $13.7\ fb^{-1}$ of data recorded by the CLEO detector at CESR, we report 
preliminary evidence for
two new charmed baryons; one decaying into $\Xi_c^{0\prime} \pi^+$ 
with the subsequent decay $\Xi_c^{0\prime} \to \Xi_c^0 \gamma$,
and its isospin partner decaying 
into $\Xi_c^{+\prime} \pi^-$ followed by $\Xi_c^{+\prime}\to\Xi_c^+\gamma$. 
We measure the following  mass differences for the two states:
$M(\Xi_c^0\gamma\pi^+)-M(\Xi_c^0)=318.4\pm1.5\pm2.9$ MeV,
and  $M(\Xi_c^+\gamma\pi^-)-M(\Xi_c^+)=323.9\pm1.4\pm3.0$ MeV. 
We interpret these new
states as the $J^P = {1 \over{2} }^-\ $ $\Xi_{c1}$ particles,
the charmed-strange analogues of the $\Lambda_{c1}^+(2593)$.

\end{abstract}
\vfill
\begin{flushleft}
.\dotfill .
\end{flushleft}
\begin{center}
Submitted to XXXth International Conference on High Energy Physics, July
2000, Osaka, Japan
\end{center}

\newpage

{
\renewcommand{\thefootnote}{\fnsymbol{footnote}}

\begin{center}
P.~Avery,$^{1}$ C.~Prescott,$^{1}$ A.~I.~Rubiera,$^{1}$
J.~Yelton,$^{1}$ J.~Zheng,$^{1}$
G.~Brandenburg,$^{2}$ A.~Ershov,$^{2}$ Y.~S.~Gao,$^{2}$
D.~Y.-J.~Kim,$^{2}$ R.~Wilson,$^{2}$
% ...
\end{center}
 
\small
\begin{center}
$^{1}${University of Florida, Gainesville, Florida 32611}\\
$^{2}${Harvard University, Cambridge, Massachusetts 02138}\\
% ...
\end{center}

\setcounter{footnote}{0}
}
\newpage

The $\Xi_c$ states consist of a combination of a charm quark, a strange quark and
an up or down quark. Each angular momentum combination of these quarks 
exists as an isospin pair. The ground states, the $\Xi_c^0$ and $\Xi_c^+$, are the only
members of the group that decay weakly, and over the past decade their masses, lifetimes,
and many of their decay modes have been measured. In 1995 and 1996 CLEO found 
evidence for a pair of excited states\cite{XICS} 
that were interpreted as the 
$J^P={3\over{2}}^+$ $\Xi_c^*$ states, and one of these observations has since been confirmed
\cite{E687}. 
In 1999, CLEO discovered\cite{XICP} 
the $\Xi_c^{\prime}$ states, which like the ground states
have $J^P={1\over{2}}^+$, but have a wave-function that is symmetric under 
interchange of the two lighter quarks, and are the charmed-strange analogs of the 
$\Sigma_c$. Also in 1999, CLEO reported the discovery\cite{XIC1} 
of a pair of states 
decaying into $\Xi_c^*\pi$ with a mass about 348 MeV above the ground states. 
These
are interpreted as the $J^P={3\over{2}}^-$ $\Xi_{c1}$ states, the analogs of the 
$\Lambda_{c1}^+(2630)$, where the numerical subscript refers to the light quark 
angular momentum. Here, using data from the
CLEO II and CLEO II.V detector configurations,  
we present the first evidence of two peaks corresponding to 
particles 
decaying to $\Xi_c^{\prime}\pi$. 
This is the expected decay mode of the $J^P={1\over{2}}^-$
$\Xi_{c1}$ states. That fact, in addition to our measured masses, 
lead us to identify
our peaks with these particles. 

The data presented here 
were taken by the CLEO detector 
operating at the Cornell Electron Storage Ring.
The sample used in this analysis corresponds to
an integrated luminosity of 13.7 $fb^{-1}$ from data
taken on the $\Upsilon(4S)$ 
resonance and in the continuum at energies just below 
the $\Upsilon(4S)$. Of this data, $4.7 fb^{-1}$ was taken with the CLEO II
configuration\cite{KUB} and the remainder with the CLEO II.V configuration\cite{HILL} 
which includes a silicon vertex detector in its charged particle measurement system.
We detected charged tracks with a cylindrical drift chamber system inside
a solenoidal magnet. Photons were detected using an electromagnetic
calorimeter consisting of 7800 cesium iodide crystals.

We first obtain large samples of reconstructed $\Xi_c^+$ and $\Xi_c^0$ 
particles, using 
their decays into $\Lambda$, $\Xi^-$, $\Omega^-$ and $\Xi^0$ hyperons
as well as kaons, pions and protons\footnote
{Charge conjugate states are implied throughout.}. 
The analysis chain for reconstructing these particles follows closely 
that presented in our previous publications \cite{XICS,XICP,XIC1}.
We fitted the invariant mass distributions for each decay mode to a sum
of a Gaussian signal function and a second order 
polynomial background. 
$\Xi_c$ candidates were defined as those combinations within $2\sigma$ of the known
mass of the $\Xi_c^+$ or $\Xi_c^0$, where $\sigma$ is the detector resolution for the
detector configuration,  
calculated mode-by-mode by a GEANT-based\cite{GEANT} Monte Carlo simulation program.
To illustrate the good statistics and
signal to noise ratio of the $\Xi_c$ signals, 
and to reduce the combinatorial background, we have placed a cut 
$x_p>0.6$, where $x_p=p/p_{max}$, $p$ is the momentum 
of the charmed baryon, $p_{max}=\sqrt{E^2_{beam}-M{^2}},$ where
$M$ is the calculated $\Xi_c$ mass. 
Table 1 details the number of signal and number of background events 
obtained from each decay mode.
The $x_p$ cut used to obtain the results in Table 1  
was not used in the final analysis, as we prefer to apply 
an $x_p$ cut only on the $\Xi_c^{\prime}\pi$ combinations.

\begin{table}[htb]
\caption{$\Xi_c^0$ and $\Xi_c^+$ Decay Mode Yields}
\begin{tabular}{cccc}
Particle&Mode&$\Xi_c$ Yield, $x_p>0.6$&Background\\
\hline
$\Xi_c^0$&&&\\
         &$\Xi^-\pi^+$            &\hfil  440  &\hfil  112 \\
         &$\Xi^-\pi^+\pi^-\pi^+$  &\hfil  106  &\hfil  133  \\
         &$\Xi^-\pi^+\pi^0$       &\hfil  357  &\hfil  377  \\
         &$\Omega^-K^+$           &\hfil   59  &\hfil   12  \\
         &$\Xi^0\pi^+\pi^-$       &\hfil  196  & \hfil 202   \\
         &$\Lambda K^-\pi^+$      &\hfil  179  &\hfil  238  \\
         &$\Lambda K^0_s$         &\hfil  106  &\hfil   80  \\
\hline
$\Xi_c^+$&&&\\
         &$\Xi^0\pi^+$            & \hfil 84  & \hfil 168  \\
         &$\Xi^0\pi^+\pi^-\pi^+$  & \hfil 216  & \hfil 460  \\
         &$\Sigma^+K^-\pi^+$      & \hfil 83  & \hfil 62  \\
         &$\Xi^-\pi^+\pi^+$       & \hfil 668  & \hfil 200  \\
         &$\Xi^0\pi^+\pi^0$       & \hfil 345  & \hfil 650  \\
         &$\Lambda K^0_s\pi^+$    & \hfil 188  & \hfil 270  \\
         &$\Lambda K^-\pi^+\pi^+$ & \hfil  38  & \hfil 78  \\
\hline
\end{tabular}
\end{table}

The $\Xi_c$ candidates defined above were then 
combined with a photon and the 
mass differences $M(\Xi_c^0\gamma)-M(\Xi_c^0)$ and 
$M(\Xi_c^+\gamma)-M(\Xi_c^+)$ were calculated. 
The transition photons were required to each have energy in excess of 100 MeV, 
to come from the part of the detector that had the best resolution 
(cos$\theta < 0.7$, where $\theta$ is the polar angle), and to have an energy
profile consistent with being due to an isolated photon. 
Any photon which, when combined
with another photon, made a combination consistent with being a $\pi^0$, was rejected.
Those combinations with calculated mass differences within 8 MeV 
($\approx 2 \sigma$) of the 
measured mass differences
for the $\Xi_c^{\prime}$ particles\cite{XICP}, were retained for further analysis.
As the $\Xi_c^{\prime}$ decays electromagnetically, its instrinsic width 
is negligible, and so the 
candidates were kinematically fit to the $\Xi_c^{\prime}$ masses 
using these measured mass differences.

We then combine these $\Xi_c^\prime$ candidates with an
appropriately charged track in the event and plot $M(\Xi_c^{\prime}\pi)-M(\Xi_c)$ 
for each isospin state. Figure 1(a) shows  $M(\Xi_c^{0\prime}\pi^+)-M(\Xi_c^0)$,
and Figure 1(b) shows  $M(\Xi_c^{+\prime}\pi^-)-M(\Xi_c^+)$, 
each with a requirement of $x_p>0.7$ on the combination. 
In both figures there is a peak at about 320 MeV, indicative of the decay 
of a $\Xi_{c1}^+$ (Figure 1a) and a $\Xi_{c1}^0$ (Figure 1b). 
We fit each of the two peaks to a sum of 
Gaussian signal function of floating width, and a
polynomial background function. 
For the $\Xi_c^{0\prime}\pi^+$ case, 
we find a signal of $18.4^{+5.6}_{-4.9}$
events with a width, $\sigma$, of $5.6\pm1.7$ MeV. 
For the $\Xi_c^{+\prime}\pi^-$ case, 
we find an excess of $14.2^{+4.6}_{-3.9}$ events 
and a width, $\sigma$, of $3.9\pm1.5$ MeV. 
The mass resolution of the detector for these decays are from our Monte Carlo
simulation program to be about 1.2 MeV in the CLEO II.V data, and around
1.4 MeV in the CLEO II data. 
This indicates the likelihood that the states have non-negligible 
intrinsic widths.
We have also fit the plots to Breit-Wigner functions 
convolved with a Gaussian resolution function.
The results of this fit are, 
$M(\Xi_c^{0\prime}\pi^+)-M(\Xi_c^0)= 318.4\pm1.5$ MeV,
and $\Gamma=7.9^{+5.7}_{-3.9}$ MeV, for Figure 1a, and 
$M(\Xi_c^{+\prime}\pi^-)-M(\Xi_c^+)= 323.9\pm1.4$ MeV,
and $\Gamma=6.5^{+4.4}_{-2.8}$ MeV for Figure 1b.
It is these fits that are superimposed on the data in Figure 1.
The results for $\Gamma$ are limited in their accuracy by the low statistics, but
indicate that it is very likely that these states have an instrinsic width of the 
order of MeV. However, we prefer to place 90\% confidence level upper limits on the width
of these states, of $\Gamma(\Xi_{c1}^+)<16\ $ MeV and $\Gamma(\Xi_{c1}^0)<13\ $ 
MeV, dominated by the statistical error.

In order to check that all the $\Xi_{c1}$ decays proceed via an 
intermediate $\Xi_c^{\prime}$, we release the cuts on $M(\Xi_c\gamma)-M(\Xi_c)$, 
select combinations within
8 MeV of our final signal peaks and plot $M(\Xi_c\gamma)-M(\Xi_c)$. 
The plots (Figures 2a and b) show signals which are
consistent with our published results for the $\Xi_c^{\prime}$ pair\cite{XIC1}.
It is clear that our data are consistent with
all the $J^P={1\over{2}}^-\ \Xi_{c1}$ decays proceeding via an intermediate $\Xi_c^{\prime}$.

When quoting our results as a  mass difference with 
respect to a ground state, our systematic
uncertainty is dominated by the uncertainty in the 
$M(\Xi_c^{\prime})-M(\Xi_c)$ mass differences. 
Alternatively, we can quote the mass difference with respect to the $\Xi_c^{\prime}$
states, and we find 
$M(\Xi_c^{+\prime}\pi^-)-M(\Xi_c^{+\prime})=216.1\pm1.4\pm1.0\ $ MeV
and 
$M(\Xi_c^{0\prime}\pi^+)-M(\Xi_c^{0\prime})=211.4\pm1.5\pm1.0\ $ MeV.
The quoted systematic uncertainties include the spread in our results obtained 
with different fitting procedures, and an estimate of the systematic uncertainty
of our mass difference scale.

The decay patterns of the $\Xi_{c1}$ states should be 
closely analogous to those of 
the $\Lambda_{c1}^+$. 
The preferred decay of the $J^P={1\over{2}}^-$ $\Xi_{c1}$ 
should be to $\Xi_c^{\prime}\pi$ because the 
spin-parity of the baryons allows this decay to
proceed via an $S$-wave decay, whereas decays to $\Xi_c^{*}$ would have to 
proceed via a $D$-wave. 
In Heavy Quark Effective Theory (HQET)\cite{IW}, where the angular momentum 
and parity of the light di-quark degrees of freedom must be
considered separately from those of the heavy quark, decays of the $\Xi_{c1}$ 
directly to ground state $\Xi_c$ baryons are not allowed.
Thus we identify our two peaks as the $J^P={1\over{2}}^-$ $\Xi_{c1}^0$ and 
$\Xi_{c1}^+$.
Also in the HQET picture, the splitting between the $J^P={1\over{2}}^-$ 
and $J^P={3\over{2}}^-$ states is expected to be similar to that 
between the two $\Lambda_{c1}^+$ states
and also the analagous splitting in charmed meson states. Combining our results
found here with our previous results on the $J^P={3\over{2}}^-\ \Xi_{c1}$ 
states\cite{XIC1}, 
and using world 
average values\cite{PDG}
for the isospin splitting of the ground state $\Xi_c$ baryons, 
we find  the splitting between the $J^P={3\over{2}}^-$
and $J^P={1\over{2}}^-$ of $24.8\pm1.8\pm3.2\ $MeV in the charged case,
and  $28.8\pm1.6\pm4.0\ $MeV for the neutral case.
These are similar to the analogous splittings in the $\Lambda_c^+$ and 
charmed meson systems. We can also measure the isospin splitting between the 
new states we have find, $M(\Xi_{c1}^0)-M(\Xi_{c1}^+)=0.0\pm2.0\pm4.5\ $MeV,
where the quoted systematic uncertainties include the systematic uncertainties
of our mass difference measurement as well as the 
uncertainty in the mass difference of the ground states. 

Although the uncertainties are large, this confirms 
the picture that the excited
states of the $\Xi_c$ have smaller isospin splittings than that of the ground states.
The $J^P={1\over{2}}\ \Xi_{c1}$ particles are the 
analogs of the $\Lambda_{c1}^+(2593)$. Although the
latter decays with very little phase space, it has been measured to have an 
instrinsic width of a few MeV. It is not surprising, therefore, 
that the $J^P={1\over{2}}^-\ \Xi_{c1}$ pair, 
which has more phase space available for two-body decays, should also appear as wide
peaks in our data.

In conclusion, we present evidence for the production of two new states. 
The first
of these states
decay into $\Xi_c^{0\prime}\pi^+$ with measured mass given by 
$M(\Xi_c^{0}\gamma\pi^+)-M(\Xi_c^0)=318.4\pm1.5\pm 2.9$ MeV.
The second state decays into $\Xi_c^{+\prime}\pi^-$ with a mass given by
$M(\Xi_c^{+}\gamma\pi^-)-M(\Xi_c^+)$ = $323.9\pm1.4\pm3.0$ MeV. 
Although we do not measure the spin or parity of these states, the observed 
decay modes, masses, 
and widths are all consistent with the new 
states being the $J^P={1\over{2}}^-$ $\Xi_{c1}^+$ and $\Xi_{c1}^0$ states, 
the charmed-strange
analogues of the $\Lambda_{c1}^+(2593)$.

\bigskip

We gratefully acknowledge the effort of the CESR staff in providing us with
excellent luminosity and running conditions.
This work was supported by 
the National Science Foundation,
the U.S. Department of Energy,
the Research Corporation,
the Natural Sciences and Engineering Research Council of Canada, 
the A.P. Sloan Foundation, 
the Swiss National Science Foundation, 
the Texas Advanced Research Program,
and the Alexander von Humboldt Stiftung.

\vfill

\begin{figure}[htb]
\noindent
\psfig{bbllx=20pt,bblly=120pt,bburx=500pt,bbury=630pt,file=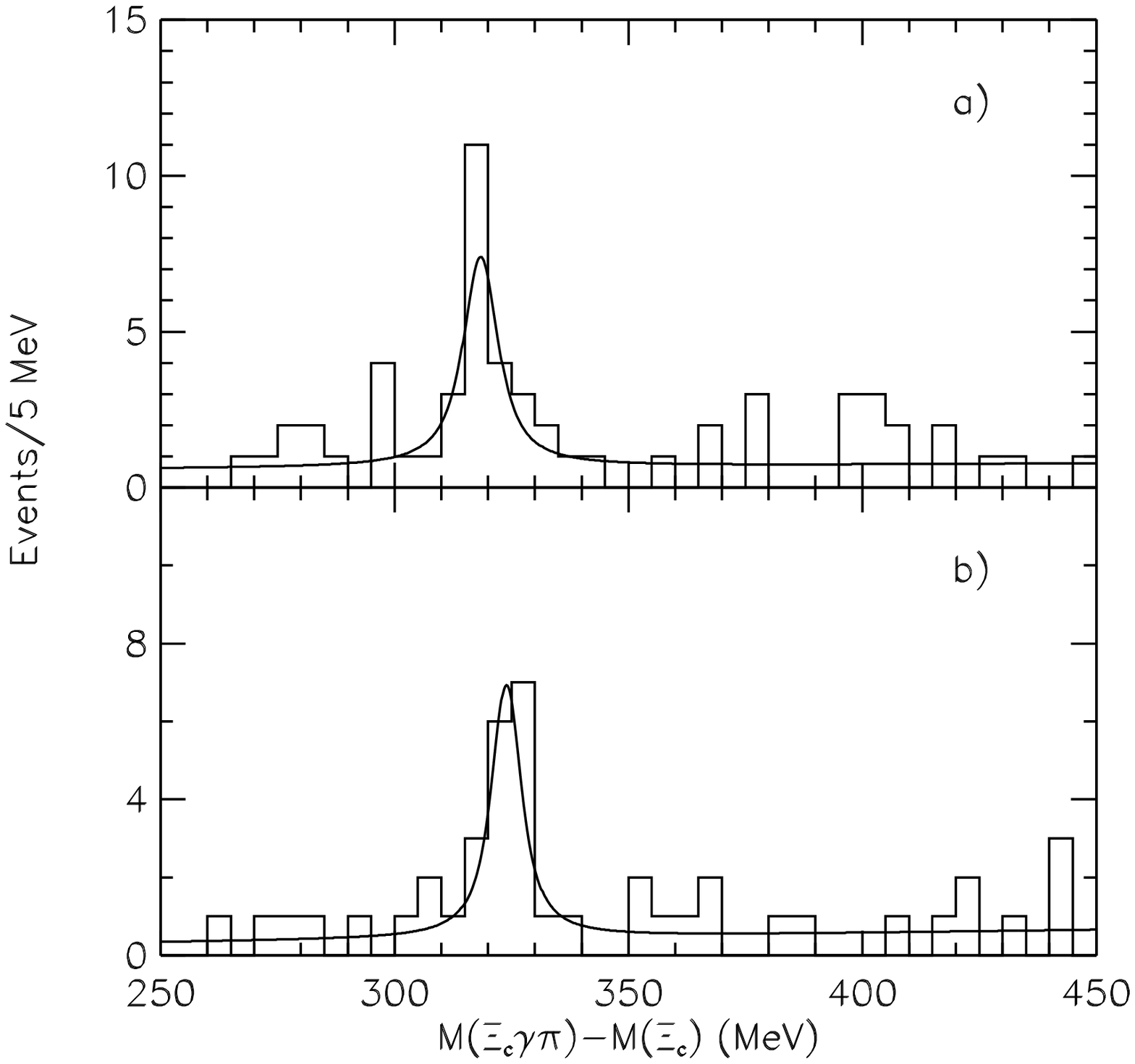,width=5.5in}
\caption[]{a) $M(\Xi_c^0\gamma\pi^+)-M(\Xi_c^0)\ $
and b) $M(\Xi_c^+\gamma\pi^-)-M(\Xi_c^+)\ $.  Each plot shows a distinct peak. Note that the
$\Xi_c^{\prime}$ mass has been fixed in these plots, so that we could equivalently
show $M(\Xi_c^{\prime}\pi)-M(\Xi_c)$ as this would just be a translation of the
horizontal axis.}
\end{figure}

\begin{figure}[htb]
\noindent
\psfig{bbllx=20pt,bblly=120pt,bburx=500pt,bbury=630pt,file=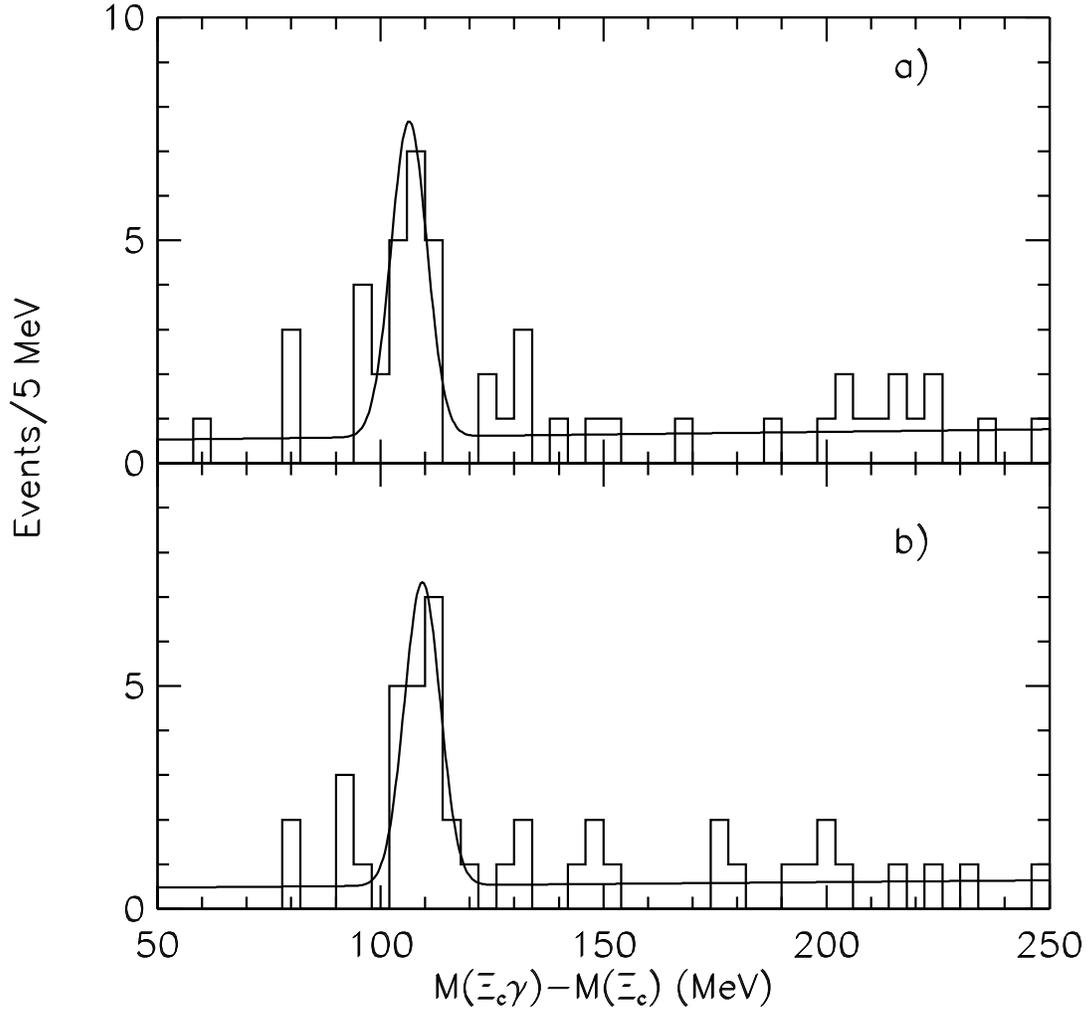,width=5.5in}
\caption[]{Cutting first on the $\Xi_{c1}$ masses, we plot $M(\Xi_c\gamma)-M(\Xi_c)$.
a) $M(\Xi_c^0\gamma)-M(\Xi_c^0)$ having cut on the $\Xi_{c1}^+$, and
b) $M(\Xi_c^+\gamma)-M(\Xi_c^+)$ having cut on the $\Xi_{c1}^0$. Each plot is fit
using a fixed width Gaussian signal function obtained from Monte Carlo simulation of 
$\Xi_c^{\prime}$ decays. The mass differences of the peaks are consistent with 
the published values for the $\Xi_c^{0\prime}$ and $\Xi_c^{+\prime}$\cite{XICP}.}

\end{figure}
\

\end{document}